\documentclass[final,3p,times,onecolumn]{elsarticle}

\usepackage{amssymb}
\usepackage{CJKutf8}
\usepackage{upgreek}

\usepackage{lineno}

\journal{NIM A}

\newcommand{\B}{\ensuremath{^{10}\mathrm{B}}}
\newcommand{\Li}{\ensuremath{^{7}\mathrm{Li}}}
\newcommand{\Am}{\ensuremath{^{241}\mathrm{Am}}}
\newcommand{\He}{\ensuremath{^{3}\mathrm{He}}}

\begin{document}
\begin{CJK*}{UTF8}{gbsn}
\begin{frontmatter}



\title{Detection of ultracold neutrons with powdered scintillator screens}


\affiliation[LANL]{%
organization=Los Alamos National Laboratory, city=Los Alamos, state=New Mexico, postcode=87545, country = USA
}%
\affiliation[NCSU]{organization=North Carolina State University, city=Raleigh, state=North Carolina, postcode = 27695, country=USA}
\affiliation[TTU]{Tennessee Technological University, Cookeville, Tennessee 38505, USA}
\affiliation[UIUC]{University of Illinois, Champaign, Illinois 61820, USA}
\affiliation[ETSU]{East Tennessee State University, Johnson City, Tennessee 37614, USA}
\affiliation[NMSU]{New Mexico State University, Las Cruces, New Mexico 88003, USA}

\author[LANL]{M. Krivos}
\ead{mkrivos@lanl.gov}
\author[LANL]{N. C. Floyd}
\author[LANL]{C. L. Morris}
\author[LANL]{Z. Tang (汤兆文)}
\author[LANL]{M. Blatnik}
\author[LANL]{S. M. Clayton}
\author[LANL]{C. B. Cude-Woods}
\author[LANL]{A. Fratangelo}
\author[TTU]{A. T. Holley}
\author[LANL]{D. E. Hooks}
\author[LANL]{T. M. Ito}
\author[UIUC]{C.-Y. Liu}
\author[LANL]{M. Makela}
\author[LANL,NMSU]{M. R. Martinez}
\author[LANL]{A. S. C. Navazo}
\author[LANL]{C. M. O'Shaughnessy}
\author[ETSU]{R. W. Pattie}
\author[LANL]{E. L. Renner}
\author[LANL]{T. A. Sandborn}
\author[LANL]{T. J. Schaub}
\author[LANL]{M. Singh}
\author[LANL]{I. L. Smythe}
\author[LANL]{F. W. Uhrich}
\author[NCSU]{N. K. Washecheck}
\author[LANL]{Z. Wang}
\author[NCSU]{A. R. Young}




\begin{abstract}
Zinc sulfide (ZnS:Ag) scintillators coated with a thin \B\ layer are widely used for ultracold neutron (UCN) detection, but their application is limited by long decay times and significant phosphorescence. We investigated two possible replacement scintillators: yttrium aluminum perovskite (YAP:Ce) and lutetium yttrium orthosilicate (LYSO:Ce). Both exhibit decay times on the order of 30–40 ns, which can help reduce dead time in high-count-rate experiments. YAP:Ce showed approximately 60\% lower phosphorescence than ZnS:Ag after 2 days and detected about 20\% more UCN. In contrast, LYSO:Ce exhibited higher phosphorescence and produced fewer UCN counts compared to both ZnS:Ag and YAP:Ce. While both tested scintillators are capable UCN detectors, YAP:Ce consistently outperformed LYSO:Ce across all measured performance metrics.
\end{abstract}





\end{frontmatter}
\end{CJK*}

\section{Introduction}

Ultracold neutrons (UCN) have kinetic energies of less than about 350 neV and can be confined gravitationally, magnetically, or by materials with a positive Fermi potential. These properties allow UCN to be used in a wide range of experiments including neutron lifetime measurements\cite{Serebrov2005, Pichlmaier2010,Pattie2018, Ezhov2018, Serebrov2018,Gonzalez2021}, neutron beta decay correlation asymmetry measurements\cite{Pattie2009,Broussard2013}, neutron electric dipole moment measurements\cite{GOLUB1994,Ito2018,Baker2006,Abel2020}, and measurements of the neutron's gravitational quantum states\cite{Nesvizhevsky2002,Pignol2007,Jenke2011}.

Traditionally, multiwire proportional chambers are employed for UCN detection by utilizing reactions with large neutron capture cross sections on \He\ or \B. For \He, the chambers are typically filled with a mixture of this high neutron capturing isotope and a carrier gas\cite{cmorris2009}. For \B, the chamber is either coated with \B\ \cite{Salvat2012,Jenke2013} or filled with a corrosive BF$_3$ gas. These detectors are easy to build and scale to required dimensions. However, one of the main issues is that the detectors require an entrance window to separate the gas from rest of the UCN volume; the window is usually made of aluminum, which has a Fermi potential barrier of about 60~neV. This means that the detectors must be installed at the end of a drop tube, so that the UCN will gain enough kinetic energy to penetrate the Fermi potential barrier of the entrance window (1-cm height difference in the Earth’s gravitational field equals approximately 1~neV of potential energy for a neutron). Such an arrangement is still subjected to UCN losses in the window itself. Another issue with these chambers is that they can be susceptible to backgrounds from thermal neutrons and high-energy charged particles.  

\begin{table*}[t]
    \centering
    \begin{tabular}{l|ccc}
                              & YAP:Ce & LYSO:Ce & ZnS:Ag \\ \hline
         Decay constant [ns]  & 28 & 40 & $>100$ \\
         Light output from $e^-$        & 25 photons/keV  & 25 photons/keV  & 52 photons/keV \\
         Peak wavelength [nm]& 370 & 410 & 450 \\
         Thickness of \B\ [nm]& 120 & 80 & 120 \\
         Area [cm$^2$]        &21.3 ± 0.5 & 20.9 ± 0.5 & 21.0 ± 0.5  \\
         Grain size [$\mu$m]  & $< 38$ & $< 38$ & - \\
         Overall thickness [$\mu$m] & 70 & 70 & - 
    \end{tabular}
    \caption{Properties of three compared scintillator screens. Decay constants and light outputs are manufacturer values. Area was measured from photographs in computer software. The grain size and thickness used in ZnS:Ag is a proprietary information of Eljen Technology.}
    \label{tab:properties}
\end{table*}

To improve the performance of UCN detection, a thin approximately 100~nm \B\ film layer coupled to zinc sulfide (ZnS:Ag) scintillator was developed\cite{WANG201530}. The ZnS:Ag scintillator was used because of its relatively low cost and high light output of 52 photons/keV. The large light output is particularly important in detecting low-energy \Li\ ions originating in the $\B(n,\alpha)\Li$ reaction. These scintillators, fabricated by Eljen technology as Eljen-440\cite{eljenZnS}, consist of two layers of scintillator powder affixed onto optical adhesive layers. The new detectors have the advantage of not being subjected to the same kind of background as the proportional chambers, since the 100 nm of \B\ will only give a 0.5\% probability for thermal neutron capture. In addition, owing to their smaller overall thickness, these scintillators experience reduced background from charged particles such as cosmic rays. The primary trade-off of ZnS:Ag is its complex decay curve, which includes long components on the order of several hundred nanoseconds and leads to increased dead time in high-count-rate experiments.

A faster scintillator would find immediate application in the UCN$\tau$(+) experiment\cite{Morris2016a, Pattie2018, Gonzalez2021}, which uses UCN for precision measurements of the neutron lifetime. UCN$\tau$ employs a custom “dagger” detector coated with ZnS:Ag scintillators, which is lowered into a UCN trap to count stored neutrons. For ZnS:Ag, the detector dead time is approximately 5~$\mu$s at UCN count rates of about 500~Hz during a 10~second dagger-lowering period. In the next-generation UCN$\tau$+ experiment, the expected count rates will be an order of magnitude higher. Under these conditions, the dead-time correction for ZnS:Ag would increase to approximately 5\%, which is unacceptable given that UCN$\tau$+ aims to measure the neutron lifetime with a precision of 0.1\%. By contrast, a faster scintillator such as YAP:Ce or LYSO:Ce would reduce the dead time by up to a factor of 70, yielding a dead-time correction of only about 0.06\%.

Another undesirable feature of ZnS:Ag is its relatively high phosphorescence. For example, in the UCN$\tau$ experiment, after exposing the detector to room light, it takes 1–2 days for the count rates to reach the background level. A scintillator with a smaller phosphorescence would allow to take more data overall.

In this work, we have investigated two possible replacement scintillators for ZnS:Ag: (1) cerium-doped yttrium aluminum perovskite (YAlO$_3$) referred to as YAP:Ce and (2) cerium-doped lutetium yttrium orthosilicate (Lu$_2$SiO$_5$) referred to as LYSO:Ce. Both scintillators have a significantly shorter decay times than ZnS:Ag, and comparable light output; see Table~\ref{tab:properties} for a summary of their properties. Section~2 describes the manufacturing process of the screens and presents results of their light output properties. Section~3 compares the UCN detection performance of YAP:Ce and LYSO:Ce with that of ZnS:Ag. 
\section{Manufacturing and light properties}
\begin{figure}
    \centering
    \includegraphics[width=0.5\linewidth]{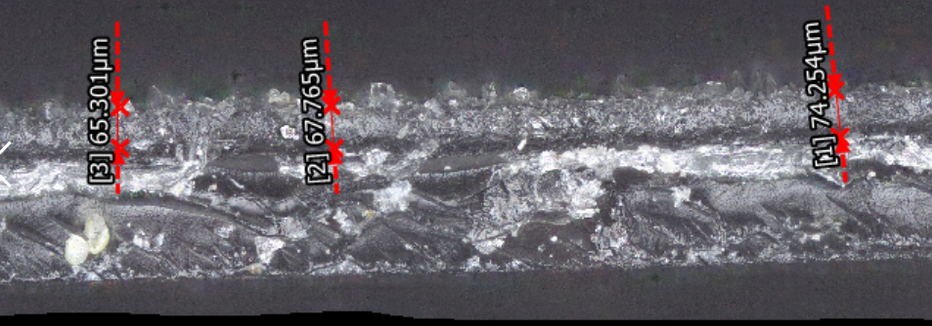}
    \caption{Cross-sectional scan of a LYSO:Ce screen. The scintillator crystal surface layer is indicated between the red arrows; the numbers adjacent to the arrows denote the layer thickness. The layer beneath the crystals corresponds to the adhesive sheet onto which the crystals are deposited.}
    \label{fig:LYSO_scan}
\end{figure}

YAP:Ce and LYSO:Ce scintillator screens were manufactured at Los Alamos National Laboratory using powder that was ground from scrap scintillator crystal material. YAP:Ce was purchased from Epic-Crystal, and LYSO:Ce from Crystal Photonics, Inc. The material was hand ground in a mortar and pestle and sized in a set of sieves. The coarser particles were reground and re-sieved. The material for the screens was selected by the finest sieve, which had a 38~$\mu$m mesh. Optical adhesive sheets from Thorlabs™ (part number OCA8146-2) were applied to ultraviolet transmitting (UVT) 25~mm acrylic sheets, the powder was rolled into the adhesive, and loose powder was brushed off.  The screens were then coated with \B\ using electron beam evaporation by Thin Films Research™. YAP:Ce and LYSO:Ce samples were cut with a razor blade and their cross-sections were imaged using a Keyence VK-X3000 laser confocal microscope (see Figure~\ref{fig:LYSO_scan}). The ZnS:Ag screen was purchased as a finished product from Eljen Technology, and its detailed properties are proprietary to the manufacturer.

\begin{figure}[!htp]
    \centering
    \includegraphics[width=0.5\linewidth]{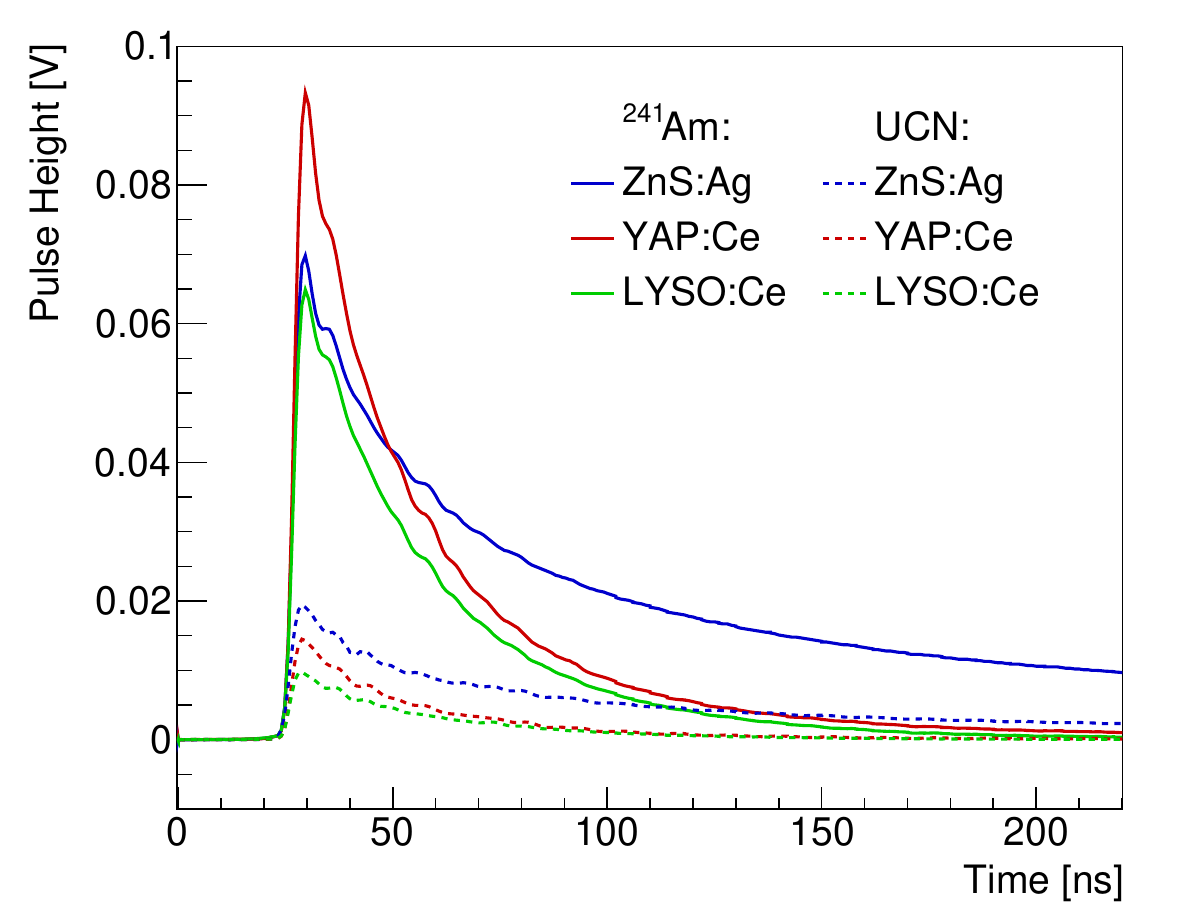}
    \caption{Pulse height of an average waveform from ZnS:Ag (blue), YAP:Ce (red), and LYSO:Ce (green). Solid lines represent scintillator activation with 5.5~MeV alphas from \Am\ while dashed lines represent UCN capture on \B.}
    \label{fig:alpha_UCN}
\end{figure}

The light output of the YAP:Ce and ZnS:Ag screens was compared using (1) 5.5 MeV alpha radiation from an \Am\ source and (2) UCN capture on the \B\ layer. In case (1), the scintillator was put in a dark box with a source directly on top of it, and in case (2) the scintillator was mounted outside a UCN test port window at the UCN source at Los Alamos National Laboratory
~\cite{Ito2018,Saunders2013}. Average waveforms for both types of measurements and all three screens are shown in Figure~\ref{fig:alpha_UCN}. The decay curve of YAP:Ce and LYSO:Ce were fitted with an exponential function, resulting in decay times of 
28~ns 
and 
32~ns, 
respectively. This result is consistent with the tabulated values for YAP:Ce from Table~\ref{tab:properties}, but is 20\% faster for LYSO:Ce. For UCN capture, the pulse heights of YAP:Ce and LYSO:Ce are approximately 75\% and 50\%, respectively, relative to ZnS:Ag. ZnS:Ag has the higher light output, as expected from the tabulated values. For 5.5 MeV alpha particles, ZnS:Ag exhibits a lower light output compared to YAP:Ce. This reduced light output may indicate a lower effective energy deposition in ZnS:Ag, which could be associated with a lower stopping power or reduced scintillation efficiency relative to YAP:Ce. The observed difference may be related to variations in the manufacturing process. In particular, differences in the concentration of optical glue, which has a lower density than the scintillator material, could reduce the effective density of the ZnS:Ag layer and thereby affect the stopping power and light output.

indicates that differences in the manufacturing process led to a screen with lower stopping power, preventing complete energy deposition of the incident higher-energy alpha particle.

\begin{figure}[!htp]
    \centering
    \includegraphics[width=0.5\linewidth]{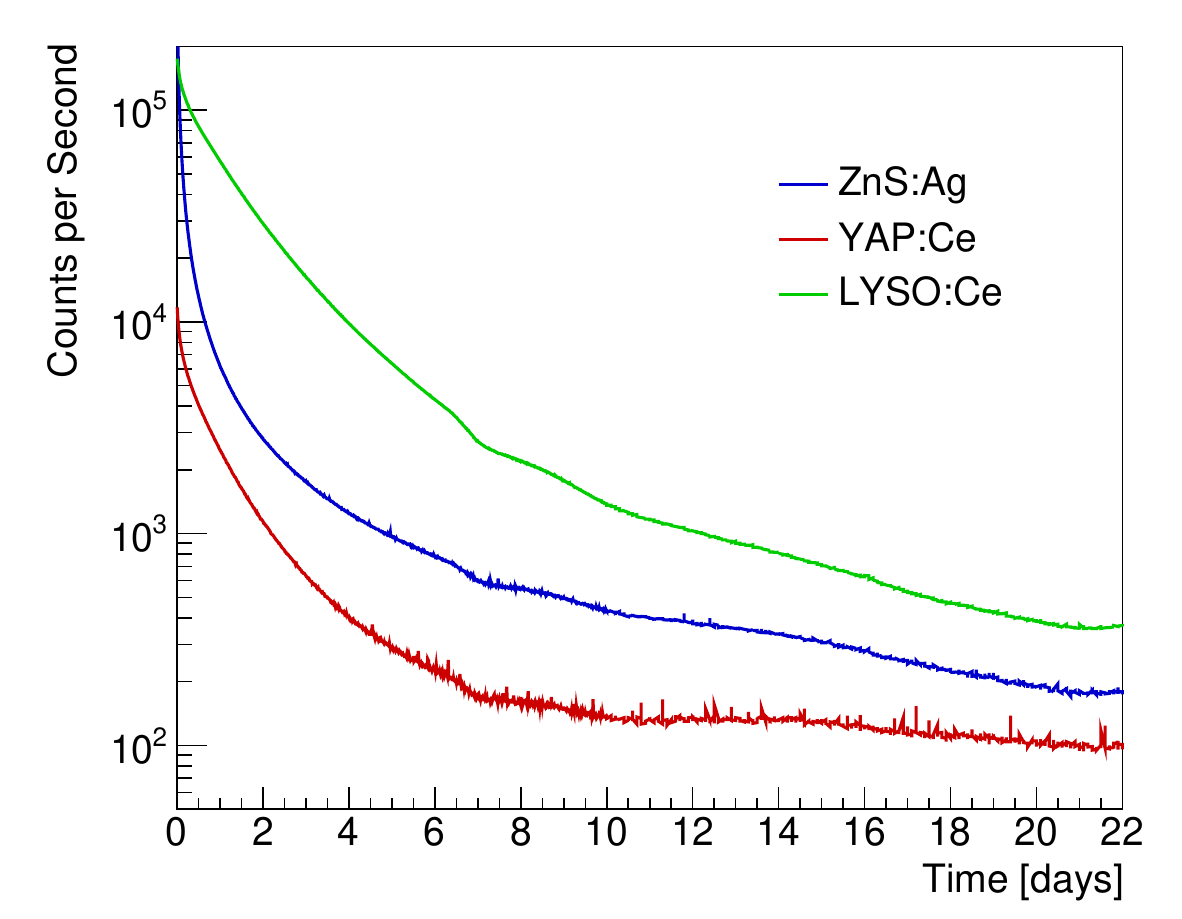}
    \caption{Phosphorescence counts from ZnS:Ag (blue), YAP:Ce (red), and LYSO:Ce (green).}
    \label{fig:phosphorescence}
\end{figure}

Phosphorescence was measured by exposing the scintillators to room light for several hours and subsequently placing them into a light-tight dark box for a period of 22 days. The resulting count rates are shown in Figure~\ref{fig:phosphorescence}. YAP:Ce shows significantly lower phosphorescence than ZnS:Ag; after 2 days, it has 60\% lower count rates. Equivalently, YAP:Ce reaches the same count-rate level in less than one day that requires two days for ZnS:Ag. Conversely, LYSO:Ce shows a much higher phosphorescence, and after 2 days exhibits over 10 times higher count rates than ZnS:Ag. The prolonged phosphorescence observed in LYSO:Ce is likely attributed to the presence of radioactive lutetium.
\section{UCN detection results}
\label{sec:ucn_detector}
\begin{figure}
    \centering
    \includegraphics[width=0.5\linewidth]{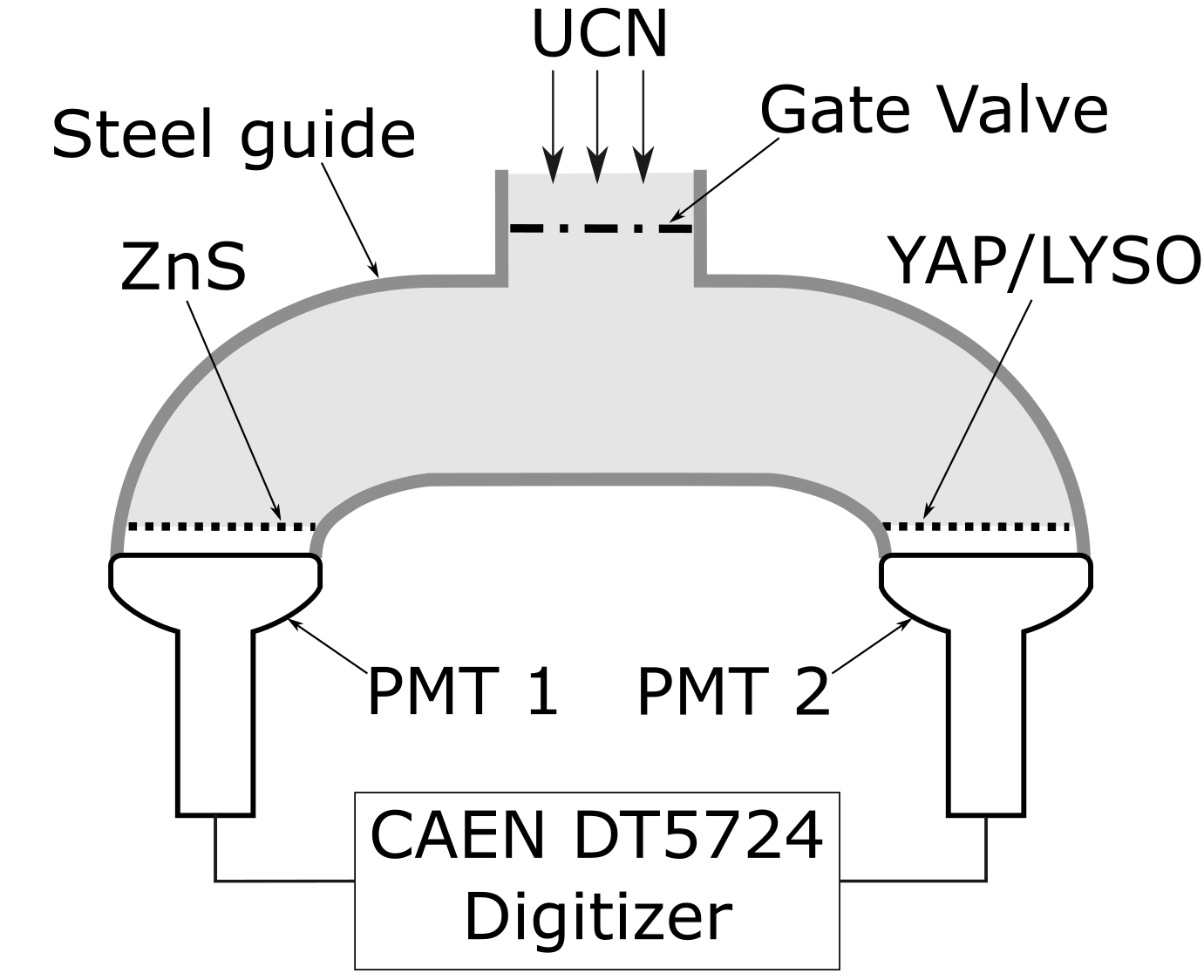}
    \caption{Experimental setup used in Sec.~\ref{sec:ucn_detector}. The gray-filled region inside the stainless-steel guide indicates where UCN are present when the gate valve is open. The dotted lines in front of the PMTs mark the positions of the tempered-glass windows with attached scintillators.}
    \label{fig:apparatus}
\end{figure}

In this section, we evaluate and compare the performance of ZnS:Ag with that of YAP:Ce and LYSO:Ce for UCN detection. Two comparative measurements were conducted: (1) between YAP:Ce and ZnS:Ag, and (2) between LYSO:Ce and ZnS:Ag. The experimental setup is shown in Fig.~\ref{fig:apparatus}. Two identical 51-mm Hamamatsu R774 photomultiplier tubes (PMTs) are symmetrically coupled to the UCN source through tempered-glass windows. A ZnS:Ag screen is placed in front of one PMT, and first YAP:Ce then LYSO:Ce screen is placed in front of the other. All scintillator screens are optically coupled to the tempered-glass using an optical adhesive to maximize light transmission.

For each scintillator pair, a 10-minute-long measurement with the UCN gate valve open (signal) and another 10-minute-long  measurement with the valve closed (background) were taken. A CAEN DT5724 100~MS/s digitizer recorded 2~$\mu$s-long waveforms from both PMTs independently. Each waveform was integrated starting at the rising edge using a 200-ns integration window. This window provides a suitable compromise for all studied scintillators: it fully captures the prompt YAP:Ce and LYSO:Ce signals while excluding the long, multi-component decay tail of ZnS:Ag, thereby integrating only the region near the signal peak. The resulting waveform integral, measured in ADC$\cdot$ns, is proportional to the scintillation light produced by heavy charged particles from the $\B(n,\alpha)\Li$ reaction and is hereafter referred to as the pulse integral.

\begin{figure}[!h]
    \centering
    \includegraphics[width=0.5\linewidth]{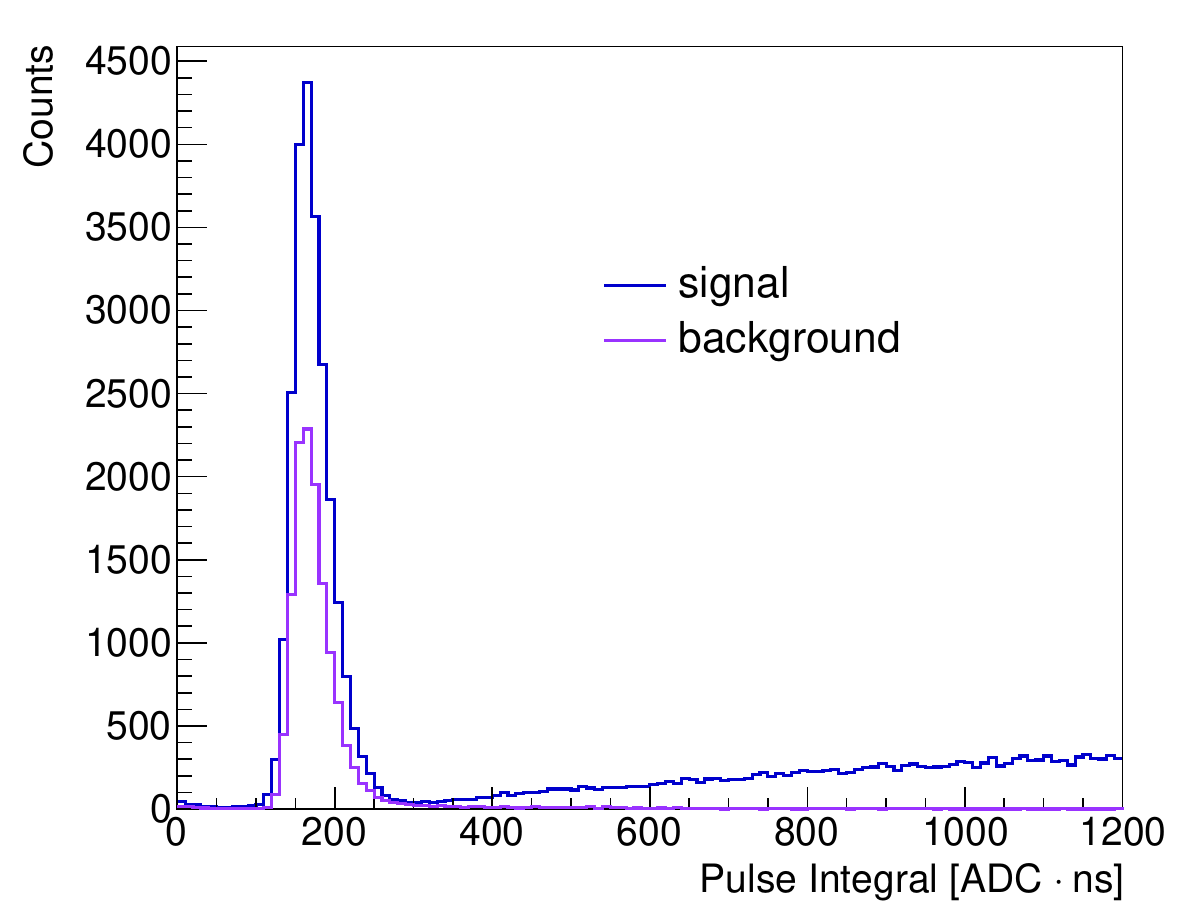}
    \caption{ZnS:Ag pulse-integral spectra for signal (blue) and background (violet) measurements. Events below 300~ADC$\cdot$ns are dominated by background and ZnS:Ag retriggering, while the increasing counts above 300~ADC$\cdot$ns correspond to genuine UCN events.}
    \label{fig:zns_bkg}
\end{figure}

\begin{figure}[!h]
    \centering
    \includegraphics[width=0.5\linewidth]{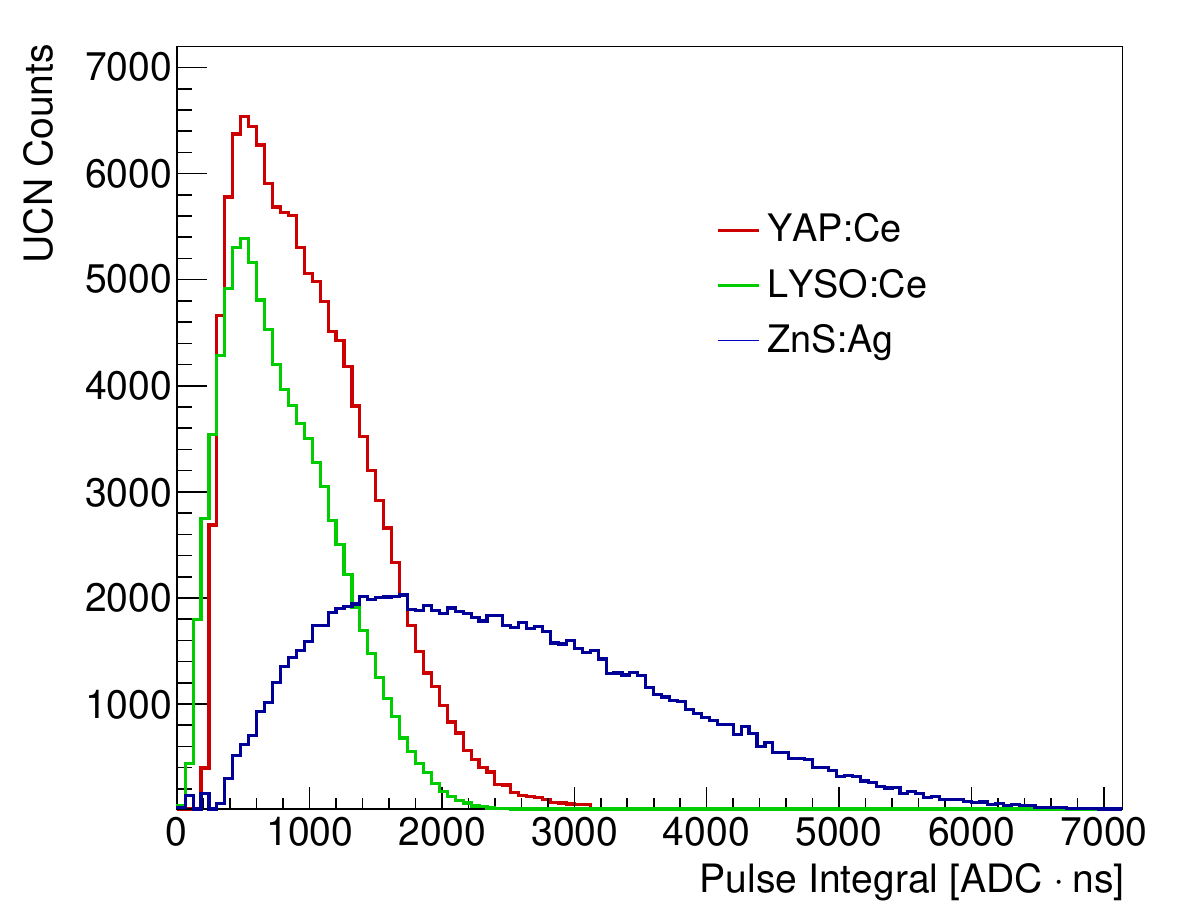}
    \caption{Pulse-integral spectra of YAP:Ce (red), LYSO:Ce (green), and ZnS:Ag (blue) after background subtraction. The LYSO:Ce spectrum was normalized using ZnS:Ag measurements. Only the ZnS:Ag spectrum acquired concurrently with YAP:Ce is shown for clarity. Mean light outputs are in the ratio $1:0.8:2.4$.}
    \label{fig:final_spectra}
\end{figure}

Background subtraction of ZnS:Ag requires a careful analysis that is explained in this paragraph. Figure~\ref{fig:zns_bkg} shows a zoomed-in view of the low values of pulse-integral spectrum for both signal and background measurement of ZnS:Ag (full pulse-integral spectrum of ZnS:Ag follows in Figure~\ref{fig:final_spectra}). 
The background spectrum consists almost entirely of a peak centered at approximately 200~ADC$\cdot$ns. For pulse-integral values above 300~ADC$\cdot$ns, the background becomes negligible and the signal region is dominated by true UCN events.
A substantial difference is observed between the number of counts in the 200~ADC$\cdot$ns peak for the signal and background measurements. This difference arises from the long ZnS:Ag decay tail, which leads to retriggering on the same UCN event (see Fig.~\ref{fig:alpha_UCN}). For this reason, experiments such as UCN$\tau$ require dead-time windows of several microseconds. While a simple post-processing dead-time window can be applied, its effectiveness is limited because the scintillation decay tail is longer than the average time between successive events. As a result, applying a sufficiently long dead time would also suppress genuine UCN counts at higher pulse-integral values.
To obtain the final background-subtracted pulse-integral spectrum, the background spectrum was scaled so that its integral matched the signal integral in the interval 0–300~ADC$\cdot$ns, where the spectrum is background-dominated. We note that counts in this region contribute less than 1\% of the total UCN counts.
\begin{figure}[!h]
    \centering
    \includegraphics[width=0.5\linewidth]{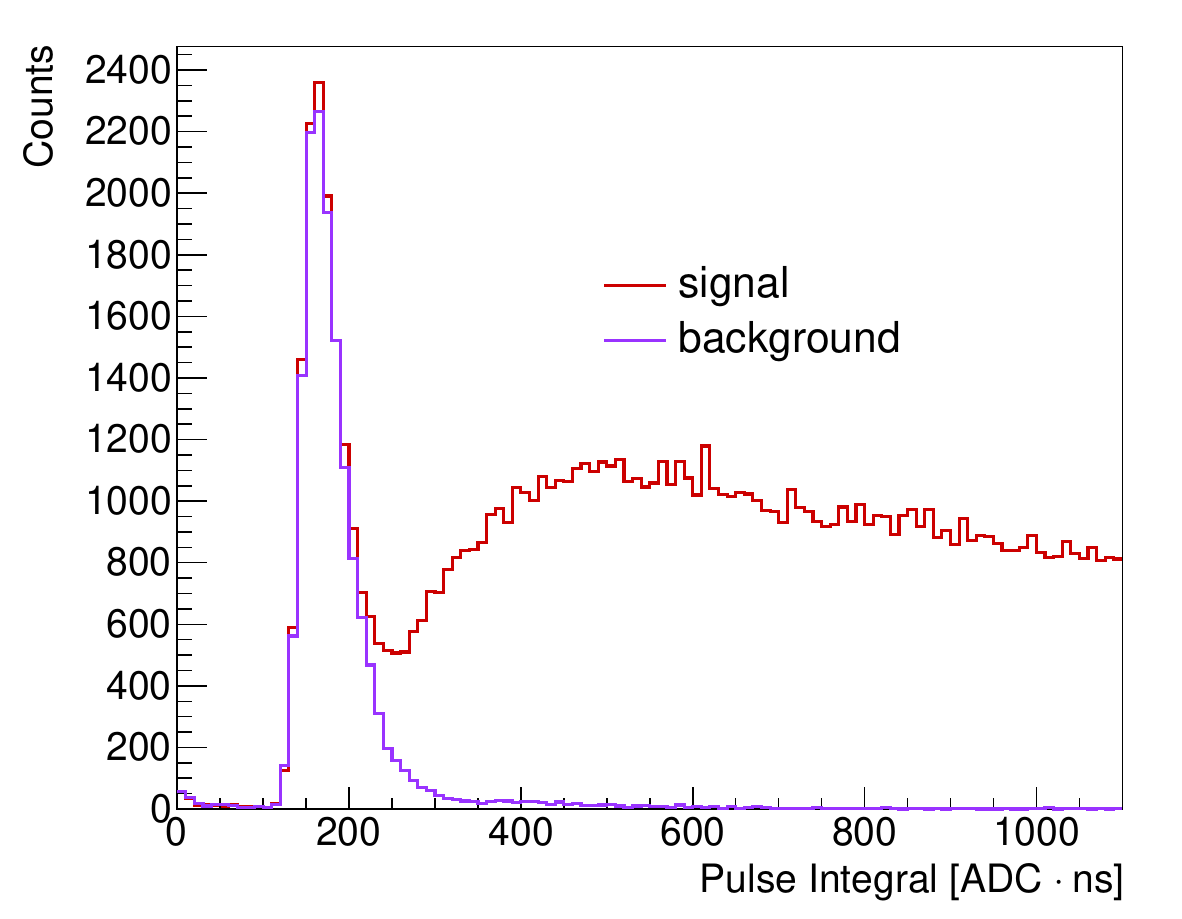}
    \caption{YAP:Ce pulse-integral spectra for signal (red) and background (violet) measurements. The background peak at approximately 200~ADC$\cdot$ns is well reproduced by the background measurement, as YAP:Ce is free of retriggering effects. The lower light output of YAP:Ce results in reduced separation between genuine UCN events and the background.}
    \label{fig:yap_bkg}
\end{figure}

YAP:Ce and LYSO:Ce are largely free of retriggering effects, and their backgrounds were subtracted without additional corrections. Figure~\ref{fig:yap_bkg} shows the low pulse-integral region of the YAP:Ce spectrum in detail. Compared to ZnS:Ag, the lower light output of YAP:Ce results in a smaller separation between the signal and background regions. This separation could be improved with a more efficient light-collection system.

\begin{table*}[h]
    \centering
    \begin{tabular}{c|cc|cc}
                 & \multicolumn{2}{c|}{YAP:Ce and ZnS:Ag measurement} & \multicolumn{2}{c}{LYSO:Ce and ZnS:Ag measurement} \\
                 \hline\hline
                            & YAP:Ce & ZnS:Ag                              & LYSO:Ce & ZnS:Ag \\
                            \hline
      Integrated UCN Counts & $127,045\pm397$   & $105,848\pm394$        & $381,859\pm691$ & $474,824\pm836$ \\
      UCN Areal Count Density [cm$^{-2}$]    & $6,002\pm142$ & $5,038\pm121$                                & $18,904\pm454$ & $22,600\pm539$   \\
      (relative uncertainty) & 2.37\% & 2.41\% & 2.40\% & 2.39\%
      \end{tabular}
    \caption{Results of UCN counting measurements from Section~\ref{sec:ucn_detector}. Integrated UCN counts uncertainties are only statistical, while UCN areal count density includes systematic uncertainty on the scintillator screen area. The relative uncertainty is shown only for UCN areal count density; it is dominated by uncertainty in the screens area measurement.}
    \label{tab:UCN_results}
\end{table*}

The background-subtracted pulse-integral spectra for the comparison between YAP:Ce and ZnS:Ag are shown in Figure~\ref{fig:final_spectra}. The lower light output of YAP:Ce results in a narrower spectrum. Total UCN counts were obtained by integrating each spectrum over all histogram bins. To account for differences in scintillator screen size, the integrated counts were normalized to the screen area, yielding UCN areal count densities. The results, summarized in Table~\ref{tab:UCN_results}, show that YAP:Ce detects approximately 20\% more UCN per square centimeter than ZnS:Ag.

\begin{figure}
    \centering
    \includegraphics[width=0.5\linewidth]{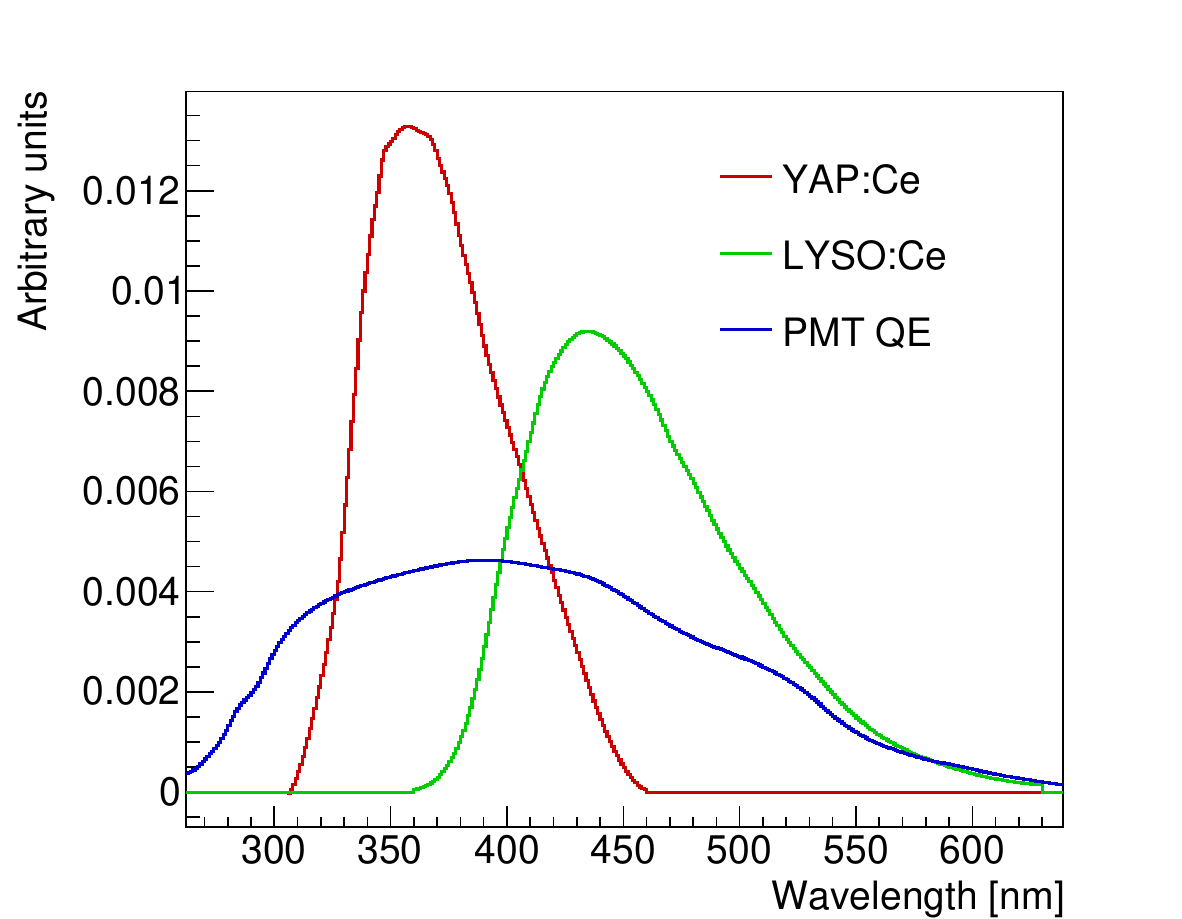}
    \caption{Emission spectra of YAP:Ce and LYSO:Ce, and quantum efficiency (QE) of the used PMT. The distributions are normalized to unit integral.}
    \label{fig:wavelenghts}
\end{figure}

In the next step, the YAP:Ce sample was replaced with LYSO:Ce and the same experimental procedure was repeated. Ten-minute-long signal and background measurements were taken, the ZnS:Ag background was estimated as in the previous case, and the LYSO:Ce background was estimated using the same method as for YAP:Ce. Due to a thinner layer of \B\ on LYSO:Ce, its UCN counts were corrected for the lower neutron capture efficiency by a factor of 1.015\footnote{For average UCN speed in our experiment, 3~m/s, and an angle of incidence of 45$^{\circ}$: $\mathrm{Exp}(n*\sigma*40~\mathrm{nm}*\sqrt{2}) = 0.13$. If we imagine 120 nm and 80 nm of \B\ coating as 3 and 2 layers of 40 nm, respectively, the correction between the two thicknesses is $(1-0.13^3)/(1-0.13^2)~=~1.015$ }.
Integration of the pulse-integral spectra for LYSO:Ce and ZnS:Ag shows that LYSO:Ce detected approximately 20\% fewer UCN per unit area than ZnS:Ag (see Table~\ref{tab:UCN_results}). Because this second measurement was performed at a different time, the spallation beam parameters and the resulting UCN density differed from those of the YAP:Ce measurement. Therefore, the corresponding ZnS:Ag measurement was used to normalize the data, allowing a direct comparison between LYSO:Ce and YAP:Ce. The normalized LYSO:Ce pulse-integral spectrum is shown in Fig.~\ref{fig:final_spectra}. 
Although YAP:Ce and LYSO:Ce both have a light yield of approximately 25 photons/keV, their detected light outputs differ due to their emission spectra. Figure~\ref{fig:wavelenghts} shows the emission spectra of both YAP:Ce and LYSO:Ce overlaid with the quantum efficiency (QE) of the PMT used in this study. Convolution of the emission spectra and the PMTs QE indicates that we should expect about 25\% more light from YAP:Ce, which agrees with our results in Figure~\ref{fig:final_spectra}. The reduced light output of LYSO:Ce likely leads to fewer detected UCN events, as some signals fall below the detection threshold.
\section{Conclusion}
We find that both investigated scintillators, yttrium aluminum perovskite (YAP:Ce) and lutetium yttrium orthosilicate (LYSO:Ce), are viable candidates for ultracold neutron (UCN) detection. Among the two, YAP:Ce demonstrates superior performance, having lower phosphorescence and a shorter decay time than both ZnS:Ag and LYSO:Ce. The UCN detection efficiency of YAP:Ce is approximately 20\% higher than that of ZnS:Ag, whereas LYSO:Ce registers about 20\% fewer counts than ZnS:Ag. The presence of naturally radioactive lutetium in LYSO:Ce results in significantly higher phosphorescence compared to both YAP:Ce and ZnS:Ag. Our results indicate a higher effective light output for YAP:Ce relative to LYSO:Ce. This difference can be explained by the distinct emission spectra of the two scintillators and their convolution with the quantum efficiency of the photomultiplier tube.
ZnS:Ag remains a suitable option for applications in which high counting rates are not expected, owing to its higher light output. 
\section{Acknowledgements}
This work was supported by the Los Alamos National Laboratory LDRD Program (Project No.  0190048 ER), National Science Foundation (Grant No. PHY1914133), and U.S. Department of Energy, Office of Nuclear Physics (Grant No. DE-FG02-97ER41042, DE-FOA-0002821). Work was performed, in part, at the Center for Integrated Nanotechnologies, an Office of Science User Facility operated for the U.S. Department of Energy (DOE) Office of Science.

\bibliography{main.bib}
\bibliographystyle{unsrt}
\end{document}